\documentclass[prd,floats,floatfix,nofootinbib]{revtex4}
\usepackage[dvips]{graphicx}
\usepackage{graphics}
\usepackage{amssymb}
\usepackage{amsmath}
\usepackage{color}

\begin{document}

\title{Optical Landscapes and High-Energy Collisions in Hairy Horndeski Gravity}

\author{Filipe Cattete Alves\footnote{filipe.ca2012@gmail.com}, Rodrigo Maier\footnote{rodrigo.maier@uerj.br}} 

\affiliation{
Departamento de F\'isica Te\'orica, Instituto de F\'isica, Universidade do Estado do Rio de Janeiro,\\
Rua S\~ao Francisco Xavier 524, Maracan\~a,\\
CEP20550-900, Rio de Janeiro, Brasil
}

\date{\today}

\begin{abstract}
We investigate the null geodesic structure, photon sphere dynamics, and high-energy particle collisions within a class of static, spherically symmetric hairy Horndeski black holes. Characterized by an invariant metric root at $r = 2M$ and a scalar hair parameter $Q$, the spacetime maps onto four distinct geometric domains dictated by the surface gravity $\kappa|_{2M}$. We derive exact analytical expressions for the photon sphere radius $r_{\text{ph}}$ and its dynamic stability criterion, showing that external circular null orbits remain dynamically unstable across non-extremal regimes. Crucially, we prove that a stable photon sphere arises exclusively in the extremal configuration ($Q = -2M$), where it coincides precisely with the degenerate horizon ($r = 2M$). We argue that this horizon-bound stable photon orbit acts as an infinitely redshifted bound state for light and ultra-relativistic particles. Finally, we analyze the Ba\~nados-Silk-West (BSW) effect for infalling timelike test particles, demonstrating that the center-of-mass energy $E_{\text{cm}}$ for critical collisions diverges as $E_{\text{cm}} \propto (r - r_h)^{-1/2}$ strictly at the extremal threshold $Q = -2M$. 
\end{abstract}

\maketitle

\section{Introduction}
\label{sec:intro}

General Relativity (GR) has successfully passed a century of observational tests, culminating in the direct detection of gravitational waves by LIGO/Virgo \cite{LIGOScientific:2016aoc} and the horizon-scale imaging of supermassive black holes by the Event Horizon Telescope (EHT) \cite{EventHorizonTelescope:2019dse, EventHorizonTelescope:2022wkp}. Despite these triumphs, GR faces deep theoretical challenges, including the dark energy problem driving cosmic acceleration, the nature of dark matter, and the persistence of spacetime singularities. These unresolved issues motivate the study of modified gravity theories as effective field theories at strong-field regimes.

Among extended theories of gravity, scalar-tensor theories represent a natural and widely studied class of modifications. Horndeski gravity \cite{Horndeski:1974wa} stands out as the most general four-dimensional scalar-tensor theory that yields second-order field equations for both the metric and the scalar field, thereby avoiding Ostrogradsky instabilities \cite{Ostrogradsky:1850fid, Motohashi:2014opa}. While classical black hole physics in vacuum GR is tightly constrained by the no-hair theorems \cite{Israel:1967wq, Carter:1971zc, Robinson:1975bv}, non-trivial couplings between the scalar field and curvature invariants in Horndeski theory allow black holes to evade these restrictions, giving rise to black holes endowed with primary or secondary scalar hair \cite{Hui:2012qt, Sotiriou:2013qea, Babichev:2013cya, Babichev:2017guv}.

Recently, Bergliaffa et al. \cite{Bergliaffa:2021diw} derived a static, spherically symmetric hairy black hole solution within a quartic subclass of Horndeski gravity. A defining feature of this spacetime is a logarithmic metric deformation governed by a scalar hair parameter $Q$. Notably, while it was originally argued that $r = 2M$ always represents an event horizon 
in the case of a nonvanishing Q, the global causal structure undergoes rich transitions across four distinct geometric domains dictated by the sign and magnitude of the surface gravity $\kappa\vert{}_{2M}$. In this paper, we reveal an overlooked aspect of this spacetime: for a specific domain of negative $Q$, $r = 2M$ actually acts as an inner Cauchy horizon rather than an event horizon.
After its derivation, the solution obtained in \cite{Bergliaffa:2021diw} has been adopted to explore strong-field phenomena. For instance, Walia et al. \cite{Walia:2021emv} extended this solution to the rotating regime using a modified Newman-Janis algorithm. Building on this, Kumar et al. \cite{Kumar:2021cyl} investigated strong gravitational lensing and shadow observables to constrain the Horndeski hair. Additionally, Jha and Rahaman \cite{Jha:2022tdl} analyzed superradiance and energy extraction, showing that the scalar parameter enhances the superradiant frequency range.

A primary observational window into strong-field gravity is provided by null geodesics and photon spheres. Unstable photon orbits govern the optical shadow cast by black holes as well as the ringdown frequency and damping rates of quasinormal modes (QNMs) in the eikonal regime \cite{Cardoso:2008bp}. Conversely, the existence of stable photon spheres -- typically associated with exotic compact objects or non-trivial horizon topologies -- can lead to stable light trapping, triggering non-linear material pile-ups and anomalous late-time power-law tails \cite{Keir:2014oka, Khoo:2016xqv}. Investigating how scalar hair alters the optical landscape is therefore crucial for constraining Horndeski gravity with upcoming observational instruments.

In addition to optical signatures, black holes serve as ultra-high-energy particle accelerators. In 2009, Ba\~nados, Silk, and West (BSW) demonstrated that the center-of-mass energy $E_{\text{cm}}$ of two test particles colliding near the horizon of an extremal Kerr black hole can formally diverge \cite{Banados:2009pr}. This BSW mechanism has since been extended to different spacetimes \cite{Zaslavskii:2010jd, Wei:2010vca}, revealing a deep connection between degenerate horizon geometries, surface gravity, and extreme particle acceleration.

In this work, we present a systematic study connecting the causal structure, photon sphere stability, and high-energy particle acceleration in hairy Horndeski spacetimes. First, we classify the global causal geometry of the Horndeski background into four distinct topological regimes governed by the hair parameter $Q$ and surface gravity $\kappa|_{2M}$. Next, by deriving exact analytical conditions for photon sphere existence and stability, we prove that a stable photon sphere arises exclusively in the extremal configuration ($Q = -2M$), serving as a concrete realization of the mechanism examined in \cite{Khoo:2016xqv}. Finally, we demonstrate that the BSW particle acceleration mechanism produces a formal divergence $E_{\text{cm}} \propto (r - r_h)^{-1/2}$ strictly at the extremal threshold $Q = -2M$, thereby establishing a direct physical link between horizon-bound stable photon orbits and extreme center-of-mass collision energies.

This paper is organized as follows. In Sec.~\ref{sec:background}, we review the hairy Horndeski spacetime of \cite{Bergliaffa:2021diw} and delineate its four geometric domains based on surface gravity. In Sec.~\ref{sec:photonsphere}, we analyze equatorial null geodesics, derive the transcendental photon sphere equation, and establish the analytical linear stability criterion. In Sec.~\ref{sec:bsw}, we study timelike particle dynamics, leading to the explicit derivation of the BSW divergence at the extremal horizon. Finally, in Sec.~\ref{sec:conclusions}, we summarize our conclusions and discuss future research directions.

\section{The Hairy Horndeski Background}
\label{sec:background}

Following the framework established in Ref.~\cite{Bergliaffa:2021diw}, we consider a static, spherically symmetric spacetime defined by the line element
\begin{equation}
    ds^2 = -f(r) dt^2 + \frac{1}{f(r)} dr^2 + r^2 \left( d\theta^2 + \sin^2\theta d\phi^2 \right),
    \label{eq:metric}
\end{equation}
where the lapse function $f(r)$ incorporates the scalar hair deformation given by a parameter $Q$, namely:
\begin{equation}
    f(r) = 1 - \frac{2M}{r} + \frac{Q}{r}\ln\left(\frac{r}{2M}\right).
    \label{eq:metric_f}
\end{equation}
In the above, $M$ denotes the asymptotic mass parameter while $Q$ acts as an effective charge tied to a unique combination of the underlying Horndeski coupling parameters. Crucially, setting $Q \rightarrow 0$
the standard Schwarzschild spacetime is restored as one should expect.

A key mathematical feature of the metric profile~\eqref{eq:metric_f} is that the coordinate radius $r = 2M$ remains an invariant root of $f(r) = 0$ for any value of the hair parameter $Q$. However, the physical role and causal status of this fixed coordinate boundary are dictated by the surface gravity $\kappa$. For a static, spherically symmetric metric of this form, the surface gravity is fundamentally defined by 
\begin{equation}
    \kappa = \frac{1}{2} \left. \frac{df}{dr} \right|_{r_{\text{root}}}.
\end{equation}
Differentiating \eqref{eq:metric_f} we then obtain
\begin{equation}
    f'(r) = \frac{1}{r^2} \left[ 2M + Q - Q\ln\left(\frac{r}{2M}\right) \right],
    \label{eq:f_prime}
\end{equation}
so that
\begin{equation}
    \kappa|_{2M} \equiv  \frac{2M + Q}{8M^2}.
    \label{eq:surface_gravity}
\end{equation}
Equation~\eqref{eq:surface_gravity} reveals that the 
causal structure of the spacetime strongly depends on the
hair parameter $Q$. While the root $r = 2M$ is invariant, its role and the broader configuration of nested horizons depend strictly on the value of $Q$, mapping out four distinct geometric domains:

\begin{itemize}
    \item { Purely Non-Extremal Domain ($Q \geq 0$):} In this domain, the surface gravity at the invariant root is strictly positive. This implies that $f(r) > 0$ for $r > 2M$ and $f(r) < 0$ for $r < 2M$. Because $\lim_{r\rightarrow 0^+} f(r) = -\infty$ when $Q \geq 0$, $r_h = 2M$ serves as a unique, standard Schwarzschild-like outer event horizon enclosing a trapped interior that terminates in a physical spacelike singularity at $r=0$. No inner horizon exists.
    
    \item { Reissner-Nordström-like Domain ($-2M < Q < 0$):} When the hair parameter becomes negative but remains above the critical threshold $-2M$, the surface gravity at the invariant root stays positive ($\kappa|_{2M} > 0$). However, because $Q < 0$, the logarithmic term dominates near the origin such that $\lim_{r\rightarrow 0^+} f(r) = +\infty$, establishing a repulsive central core. Consequently, $f(r)$ crosses from positive to negative values at $r_h = 2M$, which retains its identity as the outer event horizon. To satisfy the positive asymptotic limit at the origin, $f(r)$ must cross zero a second time at a smaller radius, $0 < r_c < 2M$. Thus, in this regime, the system possesses a nested two-horizon structure where the inner Cauchy horizon is strictly smaller than the invariant root ($r_c < 2M$).
    
    \item { Extremal Domain ($Q = -2M$):} When $Q$ reaches the critical threshold $Q = -2M$, the surface gravity vanishes identically ($\kappa|_{2M} = 0$). In this limit, the inner Cauchy horizon $r_c$ migrates outward and merges exactly with the outer event horizon at the invariant root. This configuration represents an extremal black hole characterized by a single, degenerate horizon located at $r_h = r_c = 2M$.
    
    \item { Inverted Two-Horizon Domain ($Q < -2M$):} When the hair parameter drops below the extremal threshold $-2M$, the surface gravity at the invariant root flips sign, yielding a negative value ($\kappa|_{2M} < 0$). Because the slope is negative at $r = 2M$, the lapse function must dip below zero into a trapped region immediately outside $r > 2M$. Given that asymptotic flatness requires $\lim_{r\rightarrow+\infty} f(r) = 1$, $f(r)$ must cross zero at a larger radius, $r_h > 2M$. Consequently, the outer event horizon shifts outward, while the invariant root at $r = 2M$ is demoted to the inner Cauchy horizon ($r_c = 2M$). Within this inner boundary, $f(r)$ remains strictly positive due to the repulsive core ($\lim_{r\rightarrow 0^+} f(r) = +\infty$), preserving standard coordinate signatures near the central timelike singularity.
\end{itemize}

\begin{figure}[tbp]
\includegraphics[width=8cm,height=5cm]{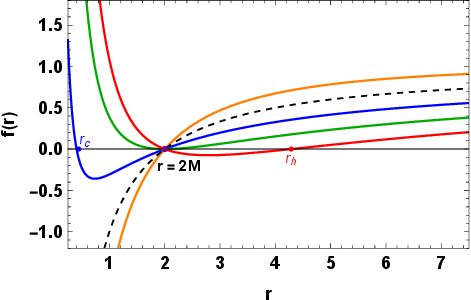}
\caption{The Horndeski lapse function $f(r)$ plotted against the radial coordinate $r$ for an asymptotic mass $M=1$ across the four distinct geometric domains. The black dot marks the invariant root at $r = 2M$. For positive hair $Q = 1$ (orange curve), $r=2M$ is a unique event horizon enclosing a Schwarzschild-like interior that plunges to $-\infty$. For the Reissner-Nordström-like domain ($Q=-1$, blue curve), $r=2M$ remains the outer event horizon, but a repulsive core emerges near the origin, forcing the creation of an inner Cauchy horizon at $r_c < 2M$ (blue dot). At the critical threshold $Q = -2M$ (green curve), these two horizons merge into a single degenerate horizon. In the inverted domain ($Q=-3$, red curve), the invariant root at $r=2M$ is demoted to the inner Cauchy horizon, while the true outer event horizon migrates outward to $r_h > 2M$ (red dot). The black dashed curve is connected to the standard Schwarzschild case in which $Q=0$.
}
\label{fig1}
\end{figure}

To visually clarify the causal transitions across these four regimes, we present the numerical profile of the lapse function $f(r)$ in Fig.~\ref{fig1} for a fixed mass scale $M=1$ across several characteristic values of the scalar hair parameter $Q$.

\section{Photon Orbits and Photon Sphere Stability}
\label{sec:photonsphere}

We now turn our attention to the optical features and null geodesic structure of the hairy Horndeski background. Specifically, we investigate the existence and stability of the photon sphere -- a compact region of closed circular null orbits -- which governs both the black hole shadow radius and the late-time quasinormal mode ringdown. Considering the spacetime given by \eqref{eq:metric_f} we focus on how the scalar hair parameter $Q$ alters standard photon orbits behavior, paying particular attention to the critical boundaries of our four geometric domains.

We begin by considering the Lagrangian for a test particle moving along a null geodesic in the equatorial plane ($\theta = \pi/2$). Utilizing the Killing vectors associated with the stationarity and spherical symmetry of the metric~\eqref{eq:metric}, we identify two conserved quantities along the affine parameter $\lambda$: the total energy $E = f(r)\dot{t}$ and the orbital angular momentum $L = r^2\dot{\phi}$, where the overdot denotes differentiation with respect to $\lambda$. The radial equation of motion can then be recast in terms of a classical effective potential $V_{\text{eff}}(r)$ via
\begin{equation}
    \dot{r}^2 = E^2 - V_{\text{eff}}(r),
    \label{eq:radial_null}
\end{equation}
where the effective potential for photons is given by
\begin{equation}
    V_{\text{eff}}(r) = \frac{L^2}{r^2} f(r) = \frac{L^2}{r^2}\left[ 1 - \frac{2M}{r} + \frac{Q}{r}\ln\left(\frac{r}{2M}\right) \right].
    \label{eq:V_eff}
\end{equation}
Circular photon orbits correspond to critical points of this effective potential, satisfying the simultaneous conditions
\begin{equation}
    V_{\text{eff}}(r_{\text{ph}}) = E^2, \quad \text{and} \quad \left. \frac{dV_{\text{eff}}}{dr} \right|_{r_{\text{ph}}} = 0.
    \label{eq:orbit_conditions}
\end{equation}
Differentiating Eq.~\eqref{eq:V_eff} with respect to $r$ yields the structural equation governing the radius of the photon sphere, $r_{\text{ph}}$:
\begin{equation}
    2 f(r_{\text{ph}}) - r_{\text{ph}} f'(r_{\text{ph}}) = 0.
    \label{eq:ph_radius_eq}
\end{equation}
Substituting the explicit expressions for the Horndeski lapse function~\eqref{eq:metric_f} and its derivative~\eqref{eq:f_prime} into Eq.~\eqref{eq:ph_radius_eq}, we find that the position of the photon sphere is determined by the transcendental relation:
\begin{equation}
\label{eq:horndeski_ph_eq}
     2 - \frac{6M}{r_{\text{ph}}} + \frac{Q}{r_{\text{ph}}} \left[ 3\ln\left(\frac{r_{\text{ph}}}{2M}\right) - 1 \right] = 0.
\end{equation}
Setting $Q \rightarrow 0$, Eq.~\ref{eq:horndeski_ph_eq} perfectly recovers the standard Schwarzschild photon sphere radius, $r_{\text{ph}}^{\text{Schw}} = 3M$ as one should expect. 
For extreme configurations it can be shown that apart an outer photon sphere located above the event horizon there is also an inner photon sphere that coincides precisely with the horizon radius. 
In the following we shall show that such inner photon sphere is connected to a stable structure
analogous to those of examined in \cite{Khoo:2016xqv}. 
In Fig. 2 we illustrate the behaviour of $r_{ph}$ with respect to the hair parameter $Q$ fixing $M=1$. 

To determine the dynamic stability of these circular null orbits, we evaluate the second derivative of the effective potential at $r = r_{\text{ph}}$. An orbit is dynamically unstable if it corresponds to a local maximum of the potential ($V_{\text{eff}}''(r_{\text{ph}}) < 0$), acting as a separator for scattered and captured light rays. Conversely, a stable photon orbit occurs at a local minimum ($V_{\text{eff}}''(r_{\text{ph}}) > 0$). A straightforward calculation yields
\begin{equation}
    V_{\text{eff}}''(r_{\text{ph}}) = \frac{L^2}{r_{\text{ph}}^2} \left[ f''(r_{\text{ph}}) -  \frac{4f'(r_{\text{ph}})}{r} +\frac{6f(r_{\text{ph}})}{r_{\text{ph}}^2} \right].
    \label{eq:potential_second_deriv}
\end{equation}
Using Eq.~\eqref{eq:ph_radius_eq} to eliminate $f(r_{\text{ph}})$, the stability criterion simplifies to the sign of the geometric profile combination:
\begin{equation}
    \text{sgn}\left[ V_{\text{eff}}''(r_{\text{ph}}) \right] = \text{sgn}\left[ r_{\text{ph}} f''(r_{\text{ph}}) - f'(r_{\text{ph}}) \right].
    \label{eq:stability_sign}
\end{equation}
Rather than evaluating the logarithmic terms numerically within Eq.~\eqref{eq:stability_sign}, we can derive a clean analytical result. By isolating the logarithmic term directly from the condition~\eqref{eq:horndeski_ph_eq}, we obtain the exact substitution:
\begin{equation}
\label{eq:log_sub}
     Q\ln\left(\frac{r_{\text{ph}}}{2M}\right) = 2M - \frac{2r_{\text{ph}}}{3} + \frac{Q}{3}.
\end{equation}
Substituting Eq.~\eqref{eq:log_sub} alongside the second derivative $f''(r)$ into Eq.~\eqref{eq:stability_sign} the transcendental stability bracket collapses into an expression that depends linearly on the photon sphere radius:
\begin{equation}
 \label{eq:stability_linear}
 \text{sgn}\left[V^{\prime\prime}_{\text{eff}}(r_{\text{ph}})\right] = \text{sgn}\left[-3Q - 2r_{\text{ph}}\right].
\end{equation}

Equation~\eqref{eq:stability_linear} provides a straightforward tool to classify the optical landscape. For the Schwarzschild baseline ($Q=0, r_{\text{ph}}=3M$), it yields $\text{sgn}\left[V^{\prime\prime}_{\text{eff}}(r_{\text{ph}})\right] < 0$, verifying the well-known instability.
\begin{figure}[tbp]
\centering
\includegraphics[width=8.5cm]{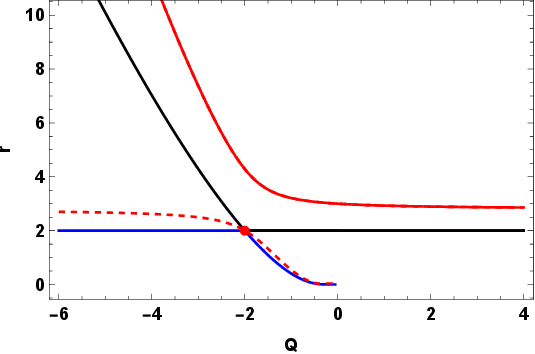} 
\caption{Radii of the photon sphere $r_{\text{ph}}$ and horizons as functions of the hair parameter $Q$ for $M=1$. The black solid curve denotes the outer event horizon $r_h$, while the blue curve tracks the inner Cauchy horizon $r_c$. Solid red curves represent physical, observable photon spheres: an outer unstable photon sphere existing across all regimes, and an inner stable photon sphere that emerges in the extreme configuration ($Q = -2M$) coinciding precisely with the degenerate horizon (red dot). The red dashed curve indicates the inner root of the transcendental equation~\eqref{eq:horndeski_ph_eq}; located within the trapped interior region between $r_h$ and $r_c$, it does not correspond to a physical photon sphere.}
\label{fig:fig2}
\end{figure}
\begin{figure}[tbp]
\centering
\includegraphics[width=8.5cm,height=5.5cm]{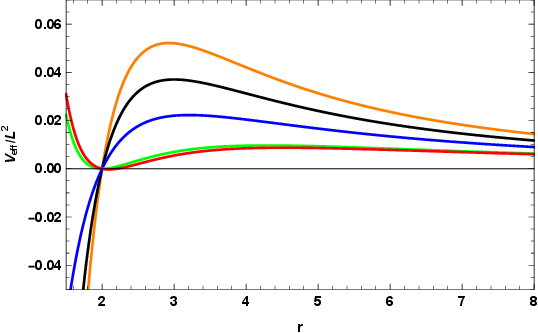}
\caption{The effective potential $V_{\text{eff}}(r)/L^2$ for equatorial null geodesics plotted against the radial coordinate $r$ for $M=1$ across different values of the hair parameter $Q$. The standard Schwarzschild baseline ($Q=0$, black curve) exhibits a single potential barrier with its maximum defining the unstable photon sphere at $r_{\text{ph}} = 3M$. In the purely non-extremal domain ($Q=1$, orange curve), positive scalar hair shifts the potential peak inward ($r_{\text{ph}} < 3M$) while increasing the barrier height. In the Reissner-Nordström-like domain ($Q=-1$, blue curve), negative hair shifts the barrier maximum outward ($r_{\text{ph}} > 3M$) while reducing its peak. At the critical threshold ($Q=-2M$, green curve), the potential peak flattens at the degenerate horizon $r=2M$, creating a local minimum ($V_{\text{eff}}''(2M) > 0$) that corresponds to a unique, stable photon sphere at the event horizon. In the inverted domain ($Q=-3$, red curve), the potential develops a multi-extremum profile with  an outer unstable maximum located deep in the exterior ($r_{\text{ph}} \simeq 7.36$).}
\label{fig3}
\end{figure}
The stability evaluation alongside the transcendental constraint unveils a subtle optical landscape, which can be visually mapped via the effective potential profiles in Fig.~\ref{fig3}. In the purely non-extremal ($Q \geq 0$) and Reissner-Nordström-like ($-2M < Q < 0$) domains, Eq.~\eqref{eq:horndeski_ph_eq} yields a single physical exterior root $r_{\text{ph}} > 2M$. Because $r_{\text{ph}} > 3|Q|/2$ holds across both regimes, the stability sign given by Eq.~\eqref{eq:stability_linear} remains strictly negative ($\text{sgn}\left[V''_{\text{eff}}\right] < 0$), confirming that external light rays are governed exclusively by standard, dynamically unstable local maxima of the potential. Even in the inverted domain ($Q < -2M$), where the potential profile develops a multi-extremum structure, the outer physical photon sphere $r_{\text{ph}} > 2M$ satisfies $2r_{\text{ph}} > -3Q$, retaining its standard unstable nature.
Crucially, a stable photon orbit arises exclusively in the extremal configuration ($Q = -2M$). In this limit, substituting $Q = -2M$ into the stability criterion~\eqref{eq:stability_linear} yields $\text{sgn}\left[V''_{\text{eff}}(2M)\right] = \text{sgn}\left[6M - 4M\right] > 0$, explicitly demonstrating that the effective potential develops a local minimum at the degenerate horizon $r_{\text{ph}} = 2M$. As shown by Khoo and Ong~\cite{Khoo:2016xqv}, photon orbits located at the event horizon of an extremal black hole exhibit distinct causal features: because $f(2M) = 0$, the effective potential vanishes identically at the minimum ($V_{\text{eff}}(2M) = 0$), meaning photons at $r_{\text{ph}} = 2M$ reside in an infinitely redshifted bound state that is marginally stable to radial perturbations. For any $Q \neq -2M$, no physical minimum with $V''_{\text{eff}} > 0$ exists outside the trapped interior. As highlighted in Ref.~\cite{Khoo:2016xqv}, such extremal stable photon spheres represent a unique topological transition boundary, which can trap null geodesics indefinitely and trigger late-time power-law tails or non-linear energy pile-ups near the horizon.

\section{The Ba\~{n}ados-Silk-West Effect and Horizon Colliders}
\label{sec:bsw}

Having established the four geometric horizon domains in Sec.~\ref{sec:background} and the dynamic stability of photon orbits in Sec.~\ref{sec:photonsphere}, we now investigate their direct physical consequences for high-energy particle collisions. Specifically, we analyze the Ba\~{n}ados-Silk-West (BSW) effect~\cite{Banados:2009pr} to determine how the Horndeski scalar hair parameter $Q$ regulates the center-of-mass energy of colliding test bodies near the horizon boundary.

Consider two uncharged test particles of identical rest mass $m_0$ dropping from rest at spatial infinity carrying distinct angular momenta $L_1$ and $L_2$. The 4-velocity of each particle $u_i^\mu = (\dot{t}_i, \dot{r}_i, 0, \dot{\phi}_i)$ moving in the equatorial plane ($\theta = \pi/2$) satisfies the normalization condition $g_{\mu\nu} u_i^\mu u_i^\nu = -1$. Utilizing the Killing symmetries of the metric~\eqref{eq:metric}, the components of the 4-velocity are given by:
\begin{equation}
    \dot{t}_i = \frac{E_i}{f(r)}, \quad \dot{\phi}_i = \frac{L_i}{r^2}, \quad \dot{r}_i = -\sqrt{E_i^2 - f(r)\left(1 + \frac{L_i^2}{r^2}\right)},
    \label{eq:4velocity_components}
\end{equation}
where the negative sign in $\dot{r}_i$ selects purely inward trajectories.

The center-of-mass energy $E_{\text{cm}}$ for a collision between these two particles at a generic radial coordinate $r$ is evaluated via the scalar product of their 4-velocities:
\begin{equation}
    \left(\frac{E_{\text{cm}}}{\sqrt{2} m_0}\right)^2 = 1 - g_{\mu\nu} u_1^\mu u_2^\nu = 1 + \frac{E_1 E_2 - \dot{r}_1 \dot{r}_2}{f(r)} - \frac{L_1 L_2}{r^2}.
    \label{eq:Ecm_general}
\end{equation}
Substituting the radial velocity profiles~\eqref{eq:4velocity_components} with $E_1 = E_2 = 1$, Eq.~\eqref{eq:Ecm_general} becomes:
\begin{equation}
    \left(\frac{E_{\text{cm}}}{\sqrt{2} m_0}\right)^2 = 1 - \frac{L_1 L_2}{r^2} + \frac{1 - \sqrt{1 - f(r)\left(1 + \frac{L_1^2}{r^2}\right)}\sqrt{1 - f(r)\left(1 + \frac{L_2^2}{r^2}\right)}}{f(r)}.
    \label{eq:Ecm_explicit_f}
\end{equation}

For a particle to physically reach a horizon located at $r = r_h$ (where $f(r_h) = 0$), its radial velocity must remain real-valued ($\dot{r}^2 \ge 0$) and its trajectory must be future-directed ($\dot{t} > 0$). Near the horizon, these physical constraints require the term under the square root in Eq.~\eqref{eq:4velocity_components} to remain non-negative, imposing a strict bound on the particle's angular momentum:
\begin{equation}
    E_i^2 - f(r)\left(1 + \frac{L_i^2}{r^2}\right) \ge 0 \implies L_i \le \frac{r E_i}{\sqrt{f(r)}}.
\end{equation}
We formally define the critical angular momentum $L_c$ as the limiting threshold value for which a particle's radial motion marginally stalls at the horizon boundary ($\lim_{r \to r_h} \dot{r} = 0$):
\begin{equation}
    L_c \equiv \lim_{r \to r_h} \frac{r E}{\sqrt{f(r)}}.
    \label{eq:Lc_def}
\end{equation}
Particles are thus naturally classified into two kinematic families, namely: 
\begin{enumerate}
    \item {Generic Subcritical Particles ($L < L_c$):} These particles possess insufficient angular momentum to overcome gravity and plunge across the horizon with a non-zero radial velocity ($\dot{r}(r_h) < 0$).
    \item {Critical Particles ($L = L_c$):} These particles carry fine-tuned angular momentum such that their radial momentum vanishes precisely at the horizon ($\dot{r}(r_h) = 0$), causing them to asymptotically hover relative to the horizon boundary.
\end{enumerate}

To determine whether $E_{\text{cm}}$ can diverge near the horizon, we evaluate Eq.~\eqref{eq:Ecm_explicit_f} as $r \to r_h$ across the geometric domains defined in Sec.~\ref{sec:background}. Performing a Taylor expansion of the lapse function $f(r)$ given by Eq.~\eqref{eq:metric_f} around $r_h$:
\begin{equation}
    f(r) = f(r_h) + f'(r_h)(r - r_h) + \frac{1}{2}f''(r_h)(r - r_h)^2 + \mathcal{O}\left((r - r_h)^3\right).
    \label{eq:f_taylor}
\end{equation}
From Eq.~\eqref{eq:surface_gravity},  we see that the linear term in Eq.~\eqref{eq:f_taylor} is directly connected to the hair parameter $Q$, that is:
\begin{eqnarray}
f^\prime(r_h)=\frac{1}{r_h^2}\Big[2M+Q-Q\ln\Big(\frac{r_h}{2M}\Big)\Big].    
\end{eqnarray}
For any hair parameter $Q \neq -2M$ we obtain $f'(r_h) > 0$. Consequently, $f(r) \propto (r - r_h)$ drops off linearly near the horizon. 

If two generic particles (with $L_1, L_2 < L_c$) collide near $r_h$, expanding the square roots in Eq.~\eqref{eq:Ecm_explicit_f} to leading order in $f(r)$ yields:
\begin{equation}
    \left(\frac{E_{\text{cm}}}{\sqrt{2}m_0}\right)^2 \approx 1 - \frac{L_1 L_2}{r_h^2} + \frac{1}{2}\left( 2 + \frac{L_1^2 + L_2^2}{r_h^2} \right) = 2 + \frac{(L_1 - L_2)^2}{2 r_h^2}.
\end{equation}
Thus, for generic particles, $E_{\text{cm}}$ approaches a strictly finite upper limit at $r \to r_h$:
\begin{equation}
    E_{\text{cm}}^2(r \to r_h) = 4m_0^2 + m_0^2 \frac{(L_1 - L_2)^2}{r_h^2},
\end{equation}
which is completely bounded regardless of the scalar hair value $Q$.

Even if one particle is critical ($L_1 = L_c$), the non-zero surface gravity $f'(r_h) > 0$ forces $L_c$ to diverge unless evaluated on an extremal horizon. Taking the limit $r \to r_h$ with linear scaling $f(r) \sim f'(r_h)(r - r_h)$ results in a $0/0$ indeterminate form where both the numerator and denominator vanish at order $\mathcal{O}(r - r_h)$. Applying L'H\^{o}pital's rule cancels the spatial fall-off, keeping $E_{\text{cm}}$ finite.

A formal divergence in $E_{\text{cm}}$ requires the critical threshold $Q = -2M$. At this boundary:
\begin{equation}
    f(2M) = 0, \quad f'(2M) = 0 \quad \text{and} \quad f''(2M) = \frac{1}{8M^3}\left[ -2M + 4M + 0 \right] = \frac{1}{4M^2}.
\end{equation}
Thus, near the degenerate horizon $r_h = 2M$, the Taylor profile~\eqref{eq:f_taylor} simplifies to a purely quadratic form dictated by $Q = -2M$:
\begin{equation}
    f(r) \approx \frac{1}{8M^2}(r - 2M)^2 + \mathcal{O}\left((r - 2M)^3\right).
    \label{eq:f_extremal_quadratic}
\end{equation}

Furthermore, substituting $Q = -2M$ and $r_h = 2M$ into the critical angular momentum condition~\eqref{eq:Lc_def} yields the finite value $L_c = 2M$. 

Now consider a collision at $r \to 2M$ between a critical particle carrying $L_1 = L_c = 2M$ and a generic particle carrying $L_2 \neq 2M$. Substituting the quadratic profile~\eqref{eq:f_extremal_quadratic} and $L_1 = 2M$ into the radial kinetic terms of Eq.~\eqref{eq:Ecm_general}, the numerator vanishes linearly as $\mathcal{O}(r - 2M)$, while the denominator vanishes quadratically as $\mathcal{O}((r - 2M)^2)$:
\begin{equation}
    \left(\frac{E_{\text{cm}}}{\sqrt{2} m_0}\right)^2 \approx \frac{\mathcal{O}(r - 2M)}{\frac{1}{8M^2}(r - 2M)^2} \propto \frac{1}{r - 2M}.
\end{equation}
Taking the explicit limit $r \to 2M$, we obtain the exact asymptotic behavior of the collision energy:
\begin{equation}
    E_{\text{cm}}^2(r \to 2M) \approx 2 m_0^2 \left[ 1 + \frac{2M - L_2}{r - 2M} \right] \xrightarrow{r \to 2M} \infty.
    \label{eq:Ecm_divergent_final}
\end{equation}

\section{Conclusions}
\label{sec:conclusions}

In this paper, we have presented an investigation into the causal structure, photon sphere stability, and high-energy particle acceleration in a class of hairy Horndeski black holes proposed in Ref.~\cite{Bergliaffa:2021diw}. By analyzing the interplay between the asymptotic mass parameter $M$ and the scalar hair charge $Q$, we mapped out the strong-field phenomenology governing both null and timelike geodesics across the four fundamental geometric domains defined by the surface gravity $\kappa|_{2M}$.

Our detailed analysis of the optical landscape revealed that for the majority of parameter space—spanning the purely non-extremal ($Q \ge 0$), Reissner-Nordström-like ($-2M < Q < 0$), and inverted ($Q < -2M$) domains—external light rays are governed exclusively by standard, dynamically unstable photon spheres ($V''_{\text{eff}}(r_{\text{ph}}) < 0$). Crucially, by evaluating the transcendental photon sphere condition alongside an exact linear stability criterion, we proved that a stable photon sphere arises exclusively in the extremal limit $Q = -2M$. In this critical threshold, the outer event horizon and inner Cauchy horizon degenerate at $r_h = r_c = 2M$, creating a horizon-bound stable photon orbit ($V''_{\text{eff}}(2M) > 0$) that vanishes in potential amplitude ($V_{\text{eff}}(2M) = 0$). This configuration acts as a concrete physical realization of the effect shown in \cite{Khoo:2016xqv}, wherein photons occupy an infinitely redshifted bound state directly at the degenerate horizon boundary.

Furthermore, we demonstrated how this topological horizon transition directly impacts timelike particle dynamics via the Bañados–Silk–West (BSW) mechanism \cite{Banados:2009pr}. Examining equatorial particle collisions near the horizon, we established that while non-extremal configurations ($Q \neq -2M$) yield finite center-of-mass energies, the extremal configuration ($Q = -2M$) catalyzes an unbound divergence $E_{\text{cm}} \propto (r - r_h)^{-1/2}$ for collisions involving a critical particle tuned to the vanishing horizon frequency. This establishes a direct physical bridge between horizon degeneracy, stable photon trapping, and infinite-energy particle acceleration in scalar-tensor gravity.

These findings open several intriguing avenues for future investigation into the strong-field observational consequences of scalar hair. A natural extension involves calculating the quasinormal mode spectrum and searching for gravitational wave echoes; because stable photon spheres are known to trap scalar and tensor modes \cite{Keir:2014oka}, a full perturbation analysis near $Q = -2M$ could reveal distinct ringdown signatures accessible to next-generation detectors. Additionally, modeling the optical shadow profile and thin-disk accretion imagery across the four geometric domains could provide direct observational constraints on the scalar hair parameter $Q$ using Event Horizon Telescope data \cite{EventHorizonTelescope:2019dse, EventHorizonTelescope:2022wkp}. In summary, this background provides a rich theoretical framework where scalar hair fundamentally reshapes horizon topology, optical trapping, and particle acceleration in modified gravity.

%%%%%%%%%%%%%%%%%%%%%%%%%%%%%%%%%%%%%%%%%%%%%%%%%%%%%%%%%%%%%%%%%%%%%%%%%%%%%%%%

\end{document}